\documentclass[%
 reprint,
 amsmath,amssymb,
 aps,
]{revtex4-2}
\usepackage{amsmath}
\usepackage{graphicx}
\usepackage{dcolumn}
\usepackage{bm}
\usepackage{footnote}
\usepackage{subfigure}%
\usepackage{hyperref}
\hypersetup{colorlinks=true, citecolor=blue, urlcolor=blue, linkcolor=blue}

\begin{document}


\title{Emergent Magnetic Monopole in Artificial Polariton Spin Ice}

\author{Junhui Cao$^{1}$}%
\email{tsao.c@mipt.ru}
\author{Alexey Kavokin$^{1,2,3}$}%
\email{a.kavokin@westlake.edu.cn}

\affiliation{$^{1}$Abrikosov Center for Theoretical Physics, Moscow Center for Advanced Studies, Moscow 141701, Russia\\
$^{2}$School of Science, Westlake University, 18 Shilongshan Road, Hangzhou 310024, Zhejiang Province, China\\
$^{3}$Department of Physics, St. Petersburg State University, University Embankment, 7/9, St. Petersburg, 199034, Russia}



\begin{abstract}
Artificial spin ice provides a versatile setting for emergent gauge fields and magnetic monopole excitations. Here we propose a driven-dissipative polariton realization of artificial spin ice, in which the circular polarization of each link mode plays the role of an Ising degree of freedom, while an auxiliary lossy vertex mode dynamically enforces a local ice-rule constraint. Adiabatic elimination of the vertex mode yields an effective spin-ice penalty, favoring the two-in two-out manifold in the steady state. We show that local polarization flips generate monopole-antimonopole defects, and that sequential flips transport these defects across the lattice while defining a Dirac string. In an extended spin-ice geometry, the vertex charges and their dynamics can be directly reconstructed from polarization-resolved real-space imaging. Our results establish polariton lattices as a controllable photonic platform for creating, manipulating, and observing emergent gauge charges in nonequilibrium spin-ice systems.
\end{abstract}

\maketitle

\paragraph*{Introduction ---}
Frustrated magnetic systems provide a fertile ground for emergent phenomena in condensed matter physics \cite{balents2010spin,batista2016frustration,diep2013frustrated,lee2002emergent,moessner2006geometrical,ramirez1994strongly,starykh2015unusual}. Among them, spin ice materials \cite{skjaervo2020advances,castelnovo2008magnetic,ramirez1999zero,zhang2013crystallites,fennell2009magnetic,ross2011quantum,petit2016observation,bramwell2001spin,bramwell2009measurement} have attracted particular attention because their local constraints lead to collective behavior analogous to gauge theories \cite{isakov2005spin,lee2012generic}. In classical spin ice, magnetic moments reside on the bonds of a lattice and obey the so-called two-in two-out ice rule \cite{10.1063/1.1749327}, meaning that two spins point into and two spins point out of every vertex. This constraint can be interpreted as a divergence-free condition on an emergent magnetic field,
\begin{equation}
\nabla \cdot S = 0,
\end{equation}
where $S$ denotes the spin field. Remarkably, violations of this rule create point-like excitations that behave as effective magnetic charges. These defects, corresponding to three-in one-out or one-in three-out configurations, have been interpreted as emergent magnetic monopoles and were experimentally observed in rare-earth pyrochlore spin ice compounds \cite{kanazawa2016critical,tokiwa2016possible,gingras2014quantum,hermele2004pyrochlore}.

The concept of emergent monopoles has inspired extensive efforts to engineer artificial spin ice systems in a variety of platforms \cite{kapaklis2012melting,kapaklis2014thermal,schiffer2021artificial,ladak2010direct,farhan2013direct}. Artificial spin ice based on lithographically fabricated magnets allows direct imaging of spin configurations and defect dynamics \cite{gartside2018realization,gartside2022reconfigurable,may2021magnetic}. Related ideas have also been explored in colloidal systems \cite{libal2012hysteresis,libal2020quenched}, superconducting circuits \cite{wang2018switchable,king2021qubit}, Rydberg atom arrays \cite{glaetzle2014quantum,shah2025quantum}. These platforms provide powerful ways to study frustration, defect propagation, and non-equilibrium dynamics beyond what is possible in natural materials. Nevertheless, many existing realizations remain limited by slow dynamics or restricted control over defect creation and manipulation.

Exciton-polariton condensates in semiconductor microcavities provide a versatile platform for exploring non-equilibrium many-body physics \cite{bloch2022non,carusotto2025exploit,cao2025exciton}. Because polaritons combine strong optical nonlinearity with direct optical accessibility, their amplitude, phase, and polarization can be measured in real time \cite{zhai2025electrically,jia2026femtosecond,shelykh2010polariton,kavokin2004quantum,utsunomiya2008observation}. Recent advances in patterned microcavities \cite{kuznetsov2023microcavity} and polariton lattices \cite{cao2025polariton} have enabled the realization of a wide variety of driven-dissipative many-body states, including topological phases \cite{klembt2018exciton,cao2022tamm}, synthetic gauge fields \cite{bieganska2021collective,chestnov2021giant}, and nonlinear collective excitations \cite{trypogeorgos2025emerging}. Moreover, exciton-polaritons in GaAs-based microcavities are intrinsically spinor particles \cite{askitopoulos2018all,PhysRevB.73.073303,PhysRevLett.97.066402}. Only bright excitons with angular-momentum projections $J_z=\pm1$ along the structure axis strongly couple to the cavity photon modes. This spinor nature is responsible for effects such as the optical spin Hall effect and the spin Meissner effect \cite{kavokin2005optical,krol2019giant}. These capabilities make polariton systems particularly promising for implementing artificial frustrated lattices with dynamically tunable interactions.

In this work we propose a scheme to realize a driven-dissipative polariton spin ice in a lattice of coupled polariton Bose-Einstein condensates \cite{hakala2018bose,bloch2022non,kavokin2022polariton}. By engineering an auxiliary lossy mode at each vertex, the system dynamically enforces a local constraint that penalizes nonzero vertex charge. Adiabatic elimination of the vertex mode leads to an effective interaction
\begin{equation}
H_{\mathrm{ice}} = \frac{U}{2}\sum_v Q_v^2 ,
\end{equation}
where
\begin{equation}
Q_v = \sum_{\ell \in v} \eta_{\ell v}\sigma_\ell
\end{equation}
represents the vertex charge defined by the polarization imbalance on the surrounding links. The lowest-penalty configurations therefore satisfy $Q_v=0$, corresponding to the two-in two-out ice rule. Violations of this constraint give rise to localized defects with $Q_v=\pm2$, which behave as emergent magnetic monopoles in the effective spin ice description. Because the polariton polarization on each link can be measured directly, the local vertex charges and their dynamics can be reconstructed experimentally. This provides a direct optical probe of monopole creation, propagation, and annihilation in a driven-dissipative lattice. Our results establish polariton lattices as a controllable platform for exploring artificial spin ice and the non-equilibrium dynamics of emergent gauge charges. The combination of strong nonlinear interactions, optical tunability, and real-time detection opens new opportunities for studying defect transport \cite{chern2013electronic}, Dirac strings \cite{jaubert2009signature,morris2009dirac,dirac1928quantum}, and collective phenomena in frustrated photonic systems.

\begin{figure}
    \centering
    \includegraphics[width=1\linewidth]{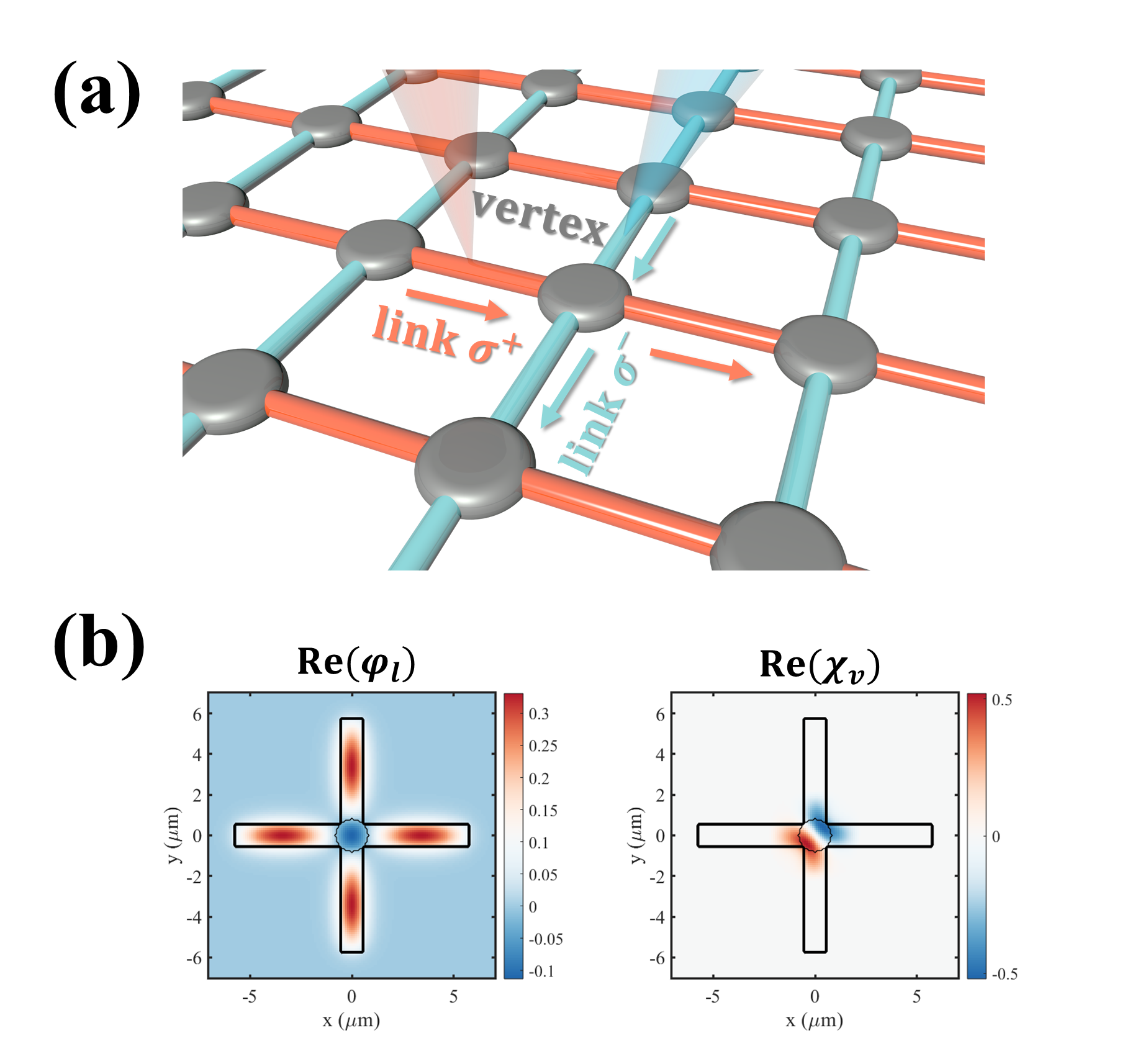}
    \caption{(a) Schematic figure of ground state of the artificial spin ice. Red and blue links represents the polariton condensates with pseudospin up and down, generated by spin-polarized pumps respectively. The vertex couples the two-in and two-out condensates. Arrows pointing towards and outwards the vertex indicate opposite lattice gauges. (b) Real part of wavefunction of the link eigenmode $\varphi_l$ and vertex eigenmode $\chi_v$. The width and length of each link are $0.6\rm\ \mu m$ and $5\rm\ \mu m$, the radius of vertex cavity is $0.8\rm\ \mu m$. Lattice gauge is designed as $(\eta_{L},\eta_{R},\eta_{D},\eta_{U})=(+1,-1,+1,-1)$. The spin-polarized pump at each link follows an elliptic Gaussian profile.}
    \label{fig:scheme}
\end{figure}

\paragraph*{Model ---}
We consider a square lattice in which each lattice link hosts a spinor polariton mode
\begin{equation}
\Psi_\ell=
\begin{pmatrix}
\psi_{\ell,+} \\
\psi_{\ell,-}
\end{pmatrix},
\end{equation}
where $\psi_{\ell,+}$ and $\psi_{\ell,-}$ denote the right and left circular polarization components (pseudospin up and down), respectively (details in Supplemental Materials (SM) Sec.I). The polarization imbalance defines an effective Ising variable on each link (same as the Stokes parameter $S_z/S_0$),
\begin{equation}
\sigma_\ell =
\frac{|\psi_{\ell,+}|^2-|\psi_{\ell,-}|^2}
     {|\psi_{\ell,+}|^2+|\psi_{\ell,-}|^2}.
\end{equation}
Fig.~\ref{fig:scheme}(a) shows the setup, while Fig.~\ref{fig:scheme}(b) shows the real part of the eigenmodes of links and vertex. We use $(\eta_{L},\eta_{R},\eta_{D},\eta_{U})=(+1,-1,+1,-1)$ as the lattice gauge for the left, right, down, up links. The in/out configuration is determined by the sign of $\sigma_\ell\eta_\ell$ on each link $\ell$. If $\rm sgn(\sigma_\ell\eta_\ell)=1$, the link is defined as ``in" (towards the vertex), while $\rm sgn(\sigma_\ell\eta_\ell)=-1$ means ``out" (outwards the vertex). A discussion on experimental realization can be found in SM Sec.IV.

Strong gain competition between the two polarization components drives the system toward states with $|\sigma_\ell|\approx 1$, effectively realizing an Ising degree of freedom on each lattice link. The dynamics of the link modes are described by
\begin{equation}
\begin{aligned}
i\dot{\psi}_{\ell,\pm}=&
\left(
\Delta - \frac{i}{2}\gamma
+\alpha |\psi_{\ell,\pm}|^2
+\beta |\psi_{\ell,\mp}|^2
\right)\psi_{\ell,\pm}\\
&+F_{\ell,\pm}
\pm g\,\eta_{\ell v} c_v ,
\end{aligned}
\label{Eq:psievo}
\end{equation}
where $F_{\ell,\pm}$ describes resonant pumping, $\gamma$ is the polariton decay rate, $\eta_{\ell v}$ is the lattice gauge constant, and $c_v$ denotes a vertex mode coupled to the links surrounding vertex $v$. The vertex mode obeys
\begin{equation}
i\dot{c}_v=
(\Delta_v-\frac{i}{2}\Gamma)c_v
+g \sum_{\ell\in v}\eta_{\ell v}(\psi_{\ell,+}-\psi_{\ell,-}),
\label{Eq:cv}
\end{equation}
where $\Gamma$ represents dissipation at the vertex. For $\Gamma \gg \gamma$, the vertex mode can be adiabatically eliminated, leading to an effective real-valued quadratic penalty functional
\begin{equation}
F_{\rm eff} = U_{\rm eff}\sum_v |\xi_v|^2 ,
\end{equation}
with
\begin{equation}
\xi_v=\sum_{\ell\in v}\eta_{\ell v}(\psi_{\ell,+}-\psi_{\ell,-}),
\qquad
U_{\rm eff}\approx\frac{2g^2}{\Gamma}.
\end{equation}
In the Ising limit, this yields the standard spin ice form
\begin{equation}
Q_v=\sum_{\ell\in v}\eta_{\ell v}\sigma_\ell ,
\end{equation}
which plays the role of a magnetic charge at vertex $v$. The driven-dissipative dynamics induces an effective steady-state penalty equivalent to a spin ice charge cost
\begin{equation}
H_{\rm ice} = \frac{U}{2}\sum_v Q_v^2 ,
\label{eq:Hice}
\end{equation}
where the prefactor $1/2$ is introduced in the conventional way for a quadratic energy penalty, such that the conjugate restoring field is simply $\partial H/\partial Q_v = U Q_v$. The coefficient $U$ therefore characterizes the energetic cost of violating the local ice-rule constraint. The detailed derivation and microscopic implementation of the important effective spin ice Hamiltonian of polaritons (Eq.~\ref{eq:Hice}) are shown in SM Sec.I-III.
Minimization of this energy requires
\begin{equation}
Q_v = 0 ,
\end{equation}
corresponding to the two-in two-out spin ice rule. Numerical simulations of the driven-dissipative dynamics confirm that the steady state of a single vertex preferentially selects configurations satisfying the ice rule, while configurations with $Q_v\neq0$ are strongly suppressed by the vertex dissipation.

\paragraph*{Emergent Magnetic Monopoles ---}

The spin ice constraint introduced above naturally gives rise to point-like topological defects that behave as emergent magnetic monopoles. In the ground state of spin ice the vertex charge $Q_v$
vanishes at every vertex, corresponding to the two-in two-out rule $Q_v=0$.
In this regime the polarization field satisfies a local divergence-free condition that is directly analogous to Gauss's law, $\nabla\cdot S = 0$.

\begin{figure}
    \centering
    \includegraphics[width=1\linewidth]{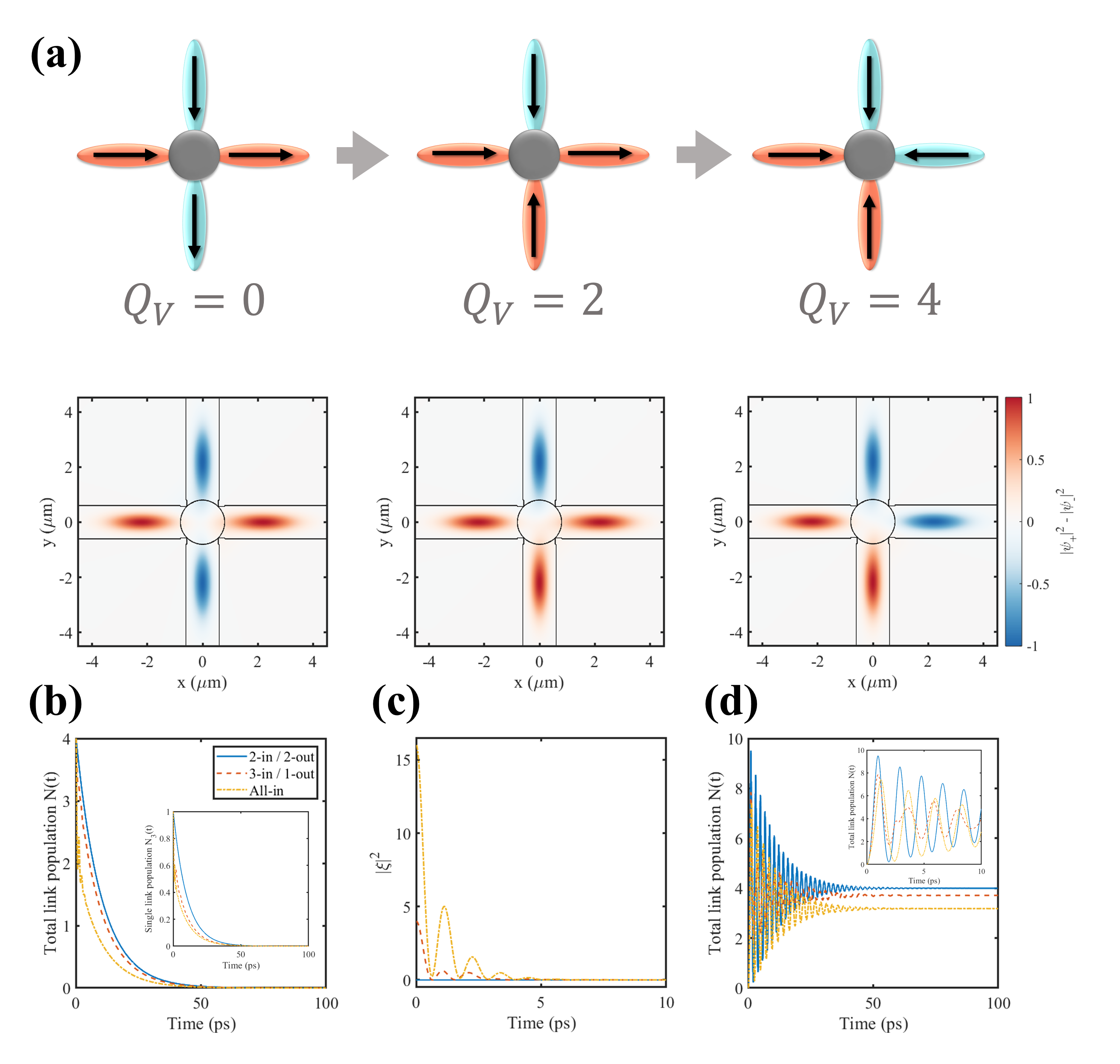}
    \caption{(a) Ground state $Q_v=0$ and excitation $Q_v=2,4$ of the artificial polariton spin ice. The black arrows indicate the in/out configurations for the links. $Q_v=0$ has 2-in 2-out configuration, while for the first excitation $Q_v=2$ it becomes 3-in 1-out and for $Q_v=4$ all arrows point to the vertex. The real space Ising variables $\sigma_\ell$ for $Q_v=0,2,4$ states are shown at the stationary state. (b) The total link population $N(t)$ for the three configurations $Q_v=0$ (blue solid curve), $Q_v=2$ (red dashed curve) and $Q_v=4$ (yellow dash-dotted curve), without external pumping. Inserted: single link population $N_3(t)$ for the down link. (c) Evolution of $|\xi|^2$ for the three configurations. (d) The total link population $N(t)$ under spin-polarized pumps with the same intensity. Inserted: a zoom-in for the time window $0-10$ ps. Parameters: $m=3\times10^{-5}m_e$ is the effective mass of polariton ($m_e$ is the mass of an electron), $\gamma=0.132 \ \rm meV$, $\Gamma=2.63 \ \rm meV$, $g=\alpha=\beta=1.58 \ \rm meV$.}
    \label{fig:excitation}
\end{figure}

A violation of the ice rule creates a nonzero vertex charge. We show this in Fig.~\ref{fig:excitation}, which summarizes the energetic selection mechanism of the artificial polariton spin ice vertex and provides direct dynamical evidence that the dissipative vertex constraint favors the ice-rule configuration. Fig.~\ref{fig:excitation}(a) illustrates three representative stationary polarization configurations of a single four-coordinated vertex, corresponding to $Q_v=0$, $Q_v=2$, and $Q_v=4$, together with their real-space distributions of the Ising-like polarization variable. In the present convention, $Q_v=0$ corresponds to the two-in two-out configuration and therefore represents the spin ice ground-state manifold. By contrast, $Q_v=2$ corresponds to a three-in one-out defect generated by flipping the polarization of the pump at the down link, while $Q_v=4$ corresponds to the fully polarized all-in configuration. The accompanying real-space $\sigma_l$ distributions clearly visualize these three distinct polarization textures.

To further quantify the stability of these configurations, in Fig.~\ref{fig:excitation}(b) we prepare the initial state in each of the three polarization sectors and let the system evolve freely without external pumping. The initial population of each link is normalized to unity, such that the total initial link population is $N(0)=4$. The subsequent decay dynamics reveal a clear hierarchy: the two-in two-out configuration ($Q_v=0$) decays slowest, the $Q_v=2$ defect decays faster, and the all-in state ($Q_v=4$) exhibits the most rapid decay. This behavior directly reflects the effective dissipative penalty induced by the lossy vertex mode. States with larger $Q_v$ generate a larger oriented polarization imbalance $\xi_v$, which enhances the vertex-induced loss channel after adiabatic elimination of the auxiliary mode. As a consequence, configurations with larger charge are dynamically suppressed, whereas the ice-rule sector survives for the longest time. This interpretation is further supported by Fig.~\ref{fig:excitation}(c), which shows the time evolution of the penalty quantity $|\xi_v|^2$ for the same three initial configurations. Consistent with the effective steady-state functional
\begin{equation}
F_{\mathrm{eff}} \propto |\xi_v|^2,
\end{equation}
the all-in state carries the largest penalty, the three-in one-out state has an intermediate penalty, and the two-in two-out state minimizes this quantity. The temporal evolution confirms that the dissipative selection rule is not merely a static classification of spin configurations, but is encoded directly in the dynamics of the coupled link-vertex system. Fig.~\ref{fig:excitation}(d) provides an even more direct signature of the role of this penalty under driven conditions. Here, the four links are driven by pumps of equal intensity, and the system is evolved from an initially empty state. If no vertex-selective penalty were present, one would expect comparable total populations for different polarization arrangements, since the external input power is identical in all cases. However, the numerical results show a strong configuration dependence: the two-in two-out state reaches a total link population close to $N = 4$, whereas the three-in one-out and all-in configurations remain at lower populations. This result demonstrates that the difference between the three sectors does not originate from unequal driving, but from their distinct effective dissipative costs. The $Q_v=0$ configuration experiences the weakest vertex-induced loss and can therefore retain nearly the full population injected by the pumps. In contrast, configurations with nonzero charge suffer stronger dissipation and are unable to sustain the same steady-state population even under identical pumping conditions.

\begin{figure*}
    \centering
    \includegraphics[width=0.9\linewidth]{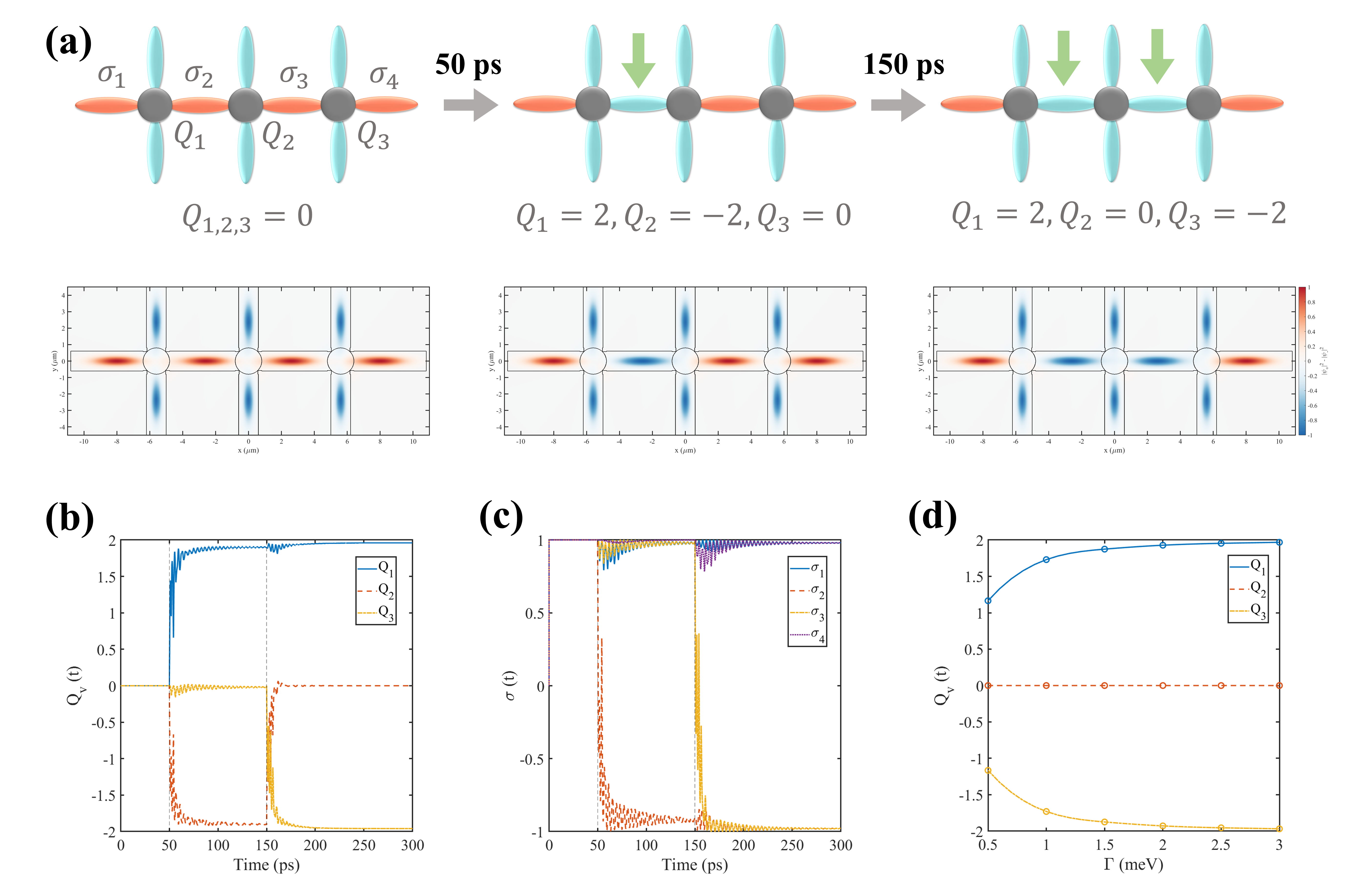}
    \caption{(a) Schematic of the protocol. The system is initially prepared in the ice-rule state with $Q_1=Q_2=Q_3=0$. At $t=50~\mathrm{ps}$, the pump polarization on link 2 is flipped, creating a monopole-antimonopole pair, $(Q_1,Q_2,Q_3)=(2,-2,0)$. At $t=150~\mathrm{ps}$, the pump polarization on link 3 is flipped, transferring the defect to the right vertex and yielding $(Q_1,Q_2,Q_3)=(2,0,-2)$. Green arrows point to the links where the pump polarization is flipped. (b) Time evolution of the vertex charges $Q_1$ (blue solid curve), $Q_2$ (red dashed curve), and $Q_3$ (yellow dash-dotted curve). (c) Time evolution of the corresponding link Ising variables $\sigma_1$ (blue solid curve), $\sigma_2$ (red dashed curve), $\sigma_3$ (yellow dash-dotted curve), and $\sigma_4$ (purple dotted curve). (d) Final vertex charges versus vertex dissipation rate $\Gamma$. Larger $\Gamma$ drives the charges closer to the ideal values $Q_v=\pm2$, consistent with the adiabatic-elimination picture and the Ising limit. Parameters are the same as in Fig.~\ref{fig:excitation} unless otherwise specified.
}
\label{fig:propagation}
\end{figure*}

Fig.~\ref{fig:propagation} extends the single-vertex physics discussed in Fig.~2 to a minimal one-dimensional spin ice lattice composed of three adjacent vertices, and demonstrates how local polarization flips can be used to controllably create, move, and read out emergent monopole defects in the driven-dissipative polariton platform. Fig.~\ref{fig:propagation}(a) presents the schematic protocol and the corresponding real-space polarization textures. From the left to right, define the vertices as $Q_{1,2,3}$ and the Ising variables on the horizontal links as $\sigma_{1,2,3,4}$, respectively. The initial state is prepared in the ice-rule manifold, such that all three vertices satisfy
\begin{equation}
Q_1=Q_2=Q_3=0.
\end{equation}
At $t=50~\mathrm{ps}$, the pump polarization on link 2 is flipped, which locally reverses the corresponding Ising variable. Since this link is shared by vertices $Q_1$ and $Q_2$, the flip changes the charges on these two neighboring vertices with opposite signs, while leaving the third vertex unaffected. As a result, a monopole-antimonopole pair is created:
\begin{equation}
(Q_1,Q_2,Q_3)=(2,-2,0).
\end{equation}
After a further 100 ps of evolution, at $t=150~\mathrm{ps}$ the pump polarization on link 3 is flipped. This second flip transfers the charge defect from the middle vertex to the right vertex, leading to
\begin{equation}
(Q_1,Q_2,Q_3)=(2,0,-2).
\end{equation}
In this way, the monopole is moved across the lattice in a controlled manner by sequential local spin flips. The string of links whose polarization has been reversed relative to the original ice-rule background constitutes the Dirac string connecting the two opposite charges.

The charge dynamics are shown explicitly in Fig.~\ref{fig:propagation}(b), which plots the time evolution of the three vertex charges $Q_1$, $Q_2$, and $Q_3$. Before the first polarization flip, all three charges remain close to zero, confirming that the initial state belongs to the two-in two-out manifold. Immediately after the first flip at $t=50~\mathrm{ps}$, $Q_1$ and $Q_2$ evolve toward approximately $+2$ and $-2$, respectively, while $Q_3$ stays near zero. This is precisely the expected signature of a monopole-antimonopole pair generated on the two vertices adjacent to the flipped link. After the second flip at $t=150~\mathrm{ps}$, the middle vertex charge $Q_2$ relaxes back toward zero, whereas $Q_3$ evolves toward $-2$, confirming the shift of the existing monopole from the second vertex to the third one. This behavior is the hallmark of defect transport in a spin ice system and provides a dynamical realization of monopole motion driven by controlled local polarization operations. Fig.~\ref{fig:propagation}(c) shows the corresponding time evolution of the link Ising variables $\sigma_1,\sigma_2,\sigma_3,\sigma_4$. Their behavior is fully consistent with the changes in the vertex charges displayed in Fig.~\ref{fig:propagation}(b). In the beginning, the four horizontal links are prepared with $\sigma^+$ polarizations, while the first abrupt change occurs on the link whose pump polarization is flipped at $50~\mathrm{ps}$. As a result, the $\sigma_2$ is flipped. Then the second change occurs on the adjacent link at $150~\mathrm{ps}$ and $\sigma_3$ is flipped. The combination of Fig.~\ref{fig:propagation}(b) and Fig.~\ref{fig:propagation}(c) thus verifies, at both the vertex and link levels, that the monopole motion follows the expected spin ice charge constraint. Fig.~\ref{fig:propagation}(d) examines the dependence of the asymptotic vertex charges on the vertex dissipation rate $\Gamma$. It can be observed that, as $\Gamma$ increases, the magnitude of the generated charges approaches more closely the ideal quantized value $|Q_v|=2$. This trend is in direct agreement with the theoretical discussion in SM Sec.II-III. In the regime of strong vertex dissipation, the auxiliary vertex mode becomes a fast variable that can be adiabatically eliminated, producing an effective penalty proportional to the square of the oriented polarization imbalance, $F_{\mathrm{eff}} \propto |\xi_v|^2$, with a strength that increases in the strongly lossy limit. At the same time, stronger dissipative selection enhances the tendency of each link to behave as an effective binary polarization degree of freedom, thereby improving the validity of the Ising-limit description of the link pseudospins. Consequently, the charge sectors become better defined and the numerically extracted values of $Q_v$ approach the ideal spin ice values $0$ and $\pm 2$.

The sequential polarization-flip protocol in Fig.~\ref{fig:propagation} provides a direct interpretation of the Dirac string in the present driven-dissipative spin ice lattice. Starting from the ice-rule manifold with $Q_v=0$ on every vertex, the first flip of an internal link creates a monopole-antimonopole pair with charges $Q=\pm2$ on the two adjacent vertices. Within the effective spin ice description, $H_{\mathrm{ice}}=\frac{U}{2}\sum_v Q_v^2$,
this operation raises the total charge cost from $0$ to
\begin{equation}
\Delta H_{\mathrm{ice}}=\frac{U}{2}\left[(+2)^2+(-2)^2\right]=4U.
\end{equation}
Importantly, subsequent flips along a connected path do not create additional charged vertices. Instead, they restore the intermediate vertex to the ice-rule sector $Q_v=0$ and transfer the nonzero charge to the next vertex. As a result, the number and magnitude of the endpoint charges remain unchanged, and the total effective penalty stays constant while the separation between the monopole and antimonopole increases. Therefore, the chain of links whose polarizations have been reversed relative to the initial ice background naturally plays the role of a Dirac string. In this sense, the Dirac string in the present model is energetically neutral in the ideal spin ice limit. Its existence is encoded in the polarization pattern, but the effective charge penalty depends only on the monopole charges at the two endpoints and not on the string length. Equivalently, the ideal string tension vanishes \cite{vedmedenko2016dynamics,stack1994string}, therefore the defect-pair energy is determined solely by monopole creation. We note that, in a realistic driven-dissipative implementation, small corrections associated with imperfect Ising polarization may generate a weak effective string tension, but the dominant behavior observed in Fig.~\ref{fig:propagation} is that monopole transport proceeds at nearly constant charge cost.

\paragraph*{Conclusion ---}
We have proposed a driven-dissipative realization of artificial spin ice in a lattice of coupled spinor polariton modes. In our scheme, the circular polarization on each link acts as an effective Ising variable, while a lossy auxiliary vertex mode induces an effective charge penalty that dynamically stabilizes the two-in two-out manifold. This mechanism provides a photonic implementation of the spin-ice constraint in terms of the degree of freedom of the polarization-resolved condensate. Our results establish polariton lattices as a promising platform for studying emergent gauge charges in a non-equilibrium setting with direct optical access, programmable driving, and real-time readout. Beyond the minimal proof-of-principle demonstrated here, this approach opens the way to exploring monopole transport and more complex frustrated photonic lattices in driven-dissipative quantum fluids.

\begin{acknowledgements}
AVK acknowledges support from Saint Petersburg State University (Research Grant No. 125022803069-4) and from the Innovation Program for Quantum Science and Technology (No. 2021ZD0302704). 
\end{acknowledgements}

\bibliography{apssamp}

\end{document}



\title{Supplemental Materials: Emergent Magnetic Monopole in Artificial Polariton Spin Ice}

\author{Junhui Cao$^{1}$}%
\email{tsao.c@mipt.ru}
\author{Alexey Kavokin$^{1,2,3}$}%
\email{a.kavokin@westlake.edu.cn}

\affiliation{$^{1}$Abrikosov Center for Theoretical Physics, Moscow Center for Advanced Studies, Moscow 141701, Russia\\
$^{2}$School of Science, Westlake University, 18 Shilongshan Road, Hangzhou 310024, Zhejiang Province, China\\
$^{3}$Department of Physics, St. Petersburg State University, University Embankment, 7/9, St. Petersburg, 199034, Russia}



\maketitle

\section{Derivation of Link and Vertex Dynamical Equations}

Let $\Phi_\pm(\mathbf r,t)$ denote the right- and left-circularly polarized polariton fields in real space. The generic driven-dissipative
spinor Gross-Pitaevskii equation is
\begin{equation}
\begin{aligned}
i\partial_t \Phi_\pm
=&
\left[
\hat H_0(\mathbf r)
-\frac{i}{2}\gamma(\mathbf r)
\right]\Phi_\pm\\
&+\alpha |\Phi_\pm|^2 \Phi_\pm
+\beta |\Phi_\mp|^2 \Phi_\pm
+F_\pm(\mathbf r,t),
\label{eq:realspace_gpe_generic}
\end{aligned}
\end{equation}
where $\hat H_0$ contains the single-particle confinement potential defining the lattice geometry, $\gamma$ is the decay rate, and $F_\pm(\mathbf r,t)$ is the coherent pump working on the links. The total field can be viewed as the superposition of localized link modes and auxiliary vertex modes, $\Phi_\pm=\Phi_{\pm,\rm link}(\mathbf r,t)+\Phi_{\pm,\rm vertex}(\mathbf r,t)$. Accordingly, we expand the real-space field of link and vertex fields as following
\begin{equation}
\begin{aligned}
\Phi_{\pm,\rm link}(\mathbf r,t)
&=
\sum_{\ell} \varphi_\ell(\mathbf r)\,\psi_{\ell,\pm}(t),
\\
\Phi_{\pm,\rm vertex}(\mathbf r,t)&=\sum_v \,u_{v,\pm}\,\chi_v(\mathbf r)c_v(t),
\label{eq:mode_expansion}
\end{aligned}
\end{equation}
where $\varphi_\ell(\mathbf r)$ is the localized orbital wavefunction of the link mode on link $\ell$, $\psi_{\ell,\pm}(t)$ is the corresponding complex mode amplitude for the two circular polarizations, $\chi_v(\mathbf r)$ is the localized spatial profile of the vertex mode; $c_v(t)$ is the complex amplitude of that vertex mode; $u_{v,\pm}$ specifies how the vertex mode couples to the two
spin components.

The oriented lattice-gauge sign $\eta_{\ell v}$ arises from the projected link-vertex coupling matrix element $g_{\ell v}$ at the junction,
\begin{equation}
g_{\ell v}
=
\int d\mathbf r\,
\varphi_\ell^*(\mathbf r)\,
\hat W(\mathbf r)\,
\chi_v(\mathbf r)\,
=g\eta_{\ell v},
\label{eq:projected_coupling_matrix_element}
\end{equation}
where $\hat W(\mathbf r)$ denotes the local coupling operator in the contact region. For simplicity, we assume $\hat W$ is a constant, hence the coupling is proportional to the overlap integral. For a vertex mode with odd spatial parity, the overlap integral acquires opposite signs on opposite links. This allows one to write the lattice gauge
\begin{equation}
(\eta_{L},\eta_{R},\eta_{D},\eta_{U})=(+1,-1,+1,-1),
\label{eq:gauge_sign_definition}
\end{equation}
as shown in Fig. 1(b). Therefore, the spatial profile $\chi_v(\mathbf r)$ determines the oriented sign pattern $\eta_{\ell v}$ through the parity of the vertex orbital, whereas the spinor coefficients $u_{v,\pm}$ enforce the relative sign between the two circular components, leading to the combination
$(\psi_{\ell,+}-\psi_{\ell,-})$ in the effective interaction Hamiltonian.

To derive Eqs.~$6-7$ in the manuscript, we assume that the auxiliary vertex mode $c_v$ couples linearly to the oriented polarization imbalance of the surrounding links. The interaction Hamiltonian is
\begin{equation}
H_{\mathrm{int}}
=
g\sum_v
\left[
c_v^\ast \sum_{\ell\in v}\eta_{\ell v}\left(\psi_{\ell,+}-\psi_{\ell,-}\right)
+\mathrm{h.c.}
\right],
\end{equation}
where $\eta_{\ell v}=\pm 1$ encodes whether link $\ell$ points into or out of vertex $v$.

The full mean-field equations are then obtained from the driven-dissipative coupled-mode dynamics
\begin{equation}
i\dot{\psi}_{\ell,\pm}
=
\frac{\partial H}{\partial \psi_{\ell,\pm}^\ast}
-\frac{i}{2}\gamma \psi_{\ell,\pm}
+F_{\ell,\pm},
\qquad
i\dot{c}_v
=
\frac{\partial H}{\partial c_v^\ast}
-\frac{i}{2}\Gamma c_v,
\end{equation}
with
\begin{equation}
\begin{aligned}
H
&=
\sum_{\ell,\sigma=\pm}\Delta |\psi_{\ell,\sigma}|^2
+\sum_v \Delta_v |c_v|^2\\
&+\sum_\ell
\left[
\frac{\alpha}{2}|\psi_{\ell,+}|^4
+\frac{\alpha}{2}|\psi_{\ell,-}|^4
+\beta |\psi_{\ell,+}|^2 |\psi_{\ell,-}|^2
\right]
+H_{\mathrm{int}}.
\end{aligned}
\end{equation}
Taking the derivatives explicitly yields
\begin{equation}
\begin{aligned}
i\dot{\psi}_{\ell,\pm}
=&
\left(
\Delta-\frac{i}{2}\gamma
+\alpha |\psi_{\ell,\pm}|^2
+\beta |\psi_{\ell,\mp}|^2
\right)\psi_{\ell,\pm}\\
&+F_{\ell,\pm}
\pm g\sum_{v\in \ell}\eta_{\ell v} c_v ,
\end{aligned}
\end{equation}
and
\begin{equation}
i\dot{c}_v
=
\left(
\Delta_v-\frac{i}{2}\Gamma
\right)c_v
+g\sum_{\ell\in v}\eta_{\ell v}\left(\psi_{\ell,+}-\psi_{\ell,-}\right).
\end{equation}
The opposite signs in the coupling to $\psi_{\ell,+}$ and $\psi_{\ell,-}$ ensure that the vertex mode couples to the polarization difference, so that after adiabatic elimination it enforces the spin ice charge constraint.

\section{Adiabatic elimination of the vertex mode}

In this section we provide a detailed derivation of the effective vertex constraint in the main text. 
Starting from the driven-dissipative equations of motion for the link and vertex modes, we adiabatically eliminate the strongly lossy vertex mode and show that this procedure generates an effective penalty proportional to the square of the local oriented polarization imbalance. 
In the Ising limit, this penalty reduces to the standard spin ice charge energy.

We now consider the regime
\begin{equation}
\Gamma \gg \gamma ,
\label{eq:fast_vertex_regime_SI}
\end{equation}
namely, the vertex mode is much more strongly damped than the link modes. 
In this limit, $c_v$ is a fast variable and can be adiabatically eliminated by setting
\begin{equation}
\dot{c}_v \approx 0.
\label{eq:adiabatic_condition_SI}
\end{equation}
Equation~(7) then gives
\begin{equation}
0=
\left(
\Delta_v-\frac{i}{2}\Gamma
\right)c_v
+
g\,\xi_v ,
\end{equation}
hence
\begin{equation}
c_v
=
-\frac{g}{\Delta_v-\frac{i}{2}\Gamma}\,\xi_v.
\label{eq:c_exact_SI}
\end{equation}
Substituting the adiabatically eliminated vertex field back into the link equation, we obtain
\begin{equation}
\pm g\eta_{\ell v}c_v
=
\mp
\frac{g^2\Delta_v}{\Delta_v^2+(\Gamma/2)^2}\eta_{\ell v}\xi_v
\mp
i\frac{g^2(\Gamma/2)}{\Delta_v^2+(\Gamma/2)^2}\eta_{\ell v}\xi_v .
\end{equation}
The first term is a coherent renormalization, whereas the second term is purely dissipative. 
In the strongly lossy regime $\Gamma\gg |\Delta_v|$, the dissipative part dominates, so that the link dynamics acquire an additional decay channel proportional to $\frac{2g^2}{\Gamma}\,\eta_{\ell v}\xi_v$. Therefore, configurations with nonzero $\xi_v$ experience enhanced loss, while the least-damped configurations satisfy $\xi_v=0$. Since the imaginary term defines a positive semi-definite quadratic form in $\xi_v$, the least-damped steady states are precisely those minimizing $|\xi_v|^2$. This motivates the real-valued steady-state penalty functional
\begin{equation}
F_{\rm eff}=U_{\rm eff}\sum_v |\xi_v|^2,
\qquad
U_{\rm eff}= \frac{g^2(\Gamma/2)}{\Delta_v^2+(\Gamma/2)^2}\approx\frac{2g^2}{\Gamma}.
\label{eq:Feff}
\end{equation}

\section{Ising limit}

To connect the continuous driven-dissipative model to spin ice, we now consider the Ising limit in which each link becomes nearly fully polarized. 
By definition,
\begin{equation}
\sigma_\ell =
\frac{|\psi_{\ell,+}|^2-|\psi_{\ell,-}|^2}
     {|\psi_{\ell,+}|^2+|\psi_{\ell,-}|^2},
\qquad
-1 \le \sigma_\ell \le 1.
\end{equation}
Strong gain competition between the two circular polarizations drives the system toward
\begin{equation}
|\sigma_\ell| \approx 1,
\end{equation}
so that each link effectively realizes a binary variable
\begin{equation}
\sigma_\ell = \pm 1.
\label{eq:Ising_binary_SI}
\end{equation}
Let the total intensity on each link be
\begin{equation}
n_\ell = |\psi_{\ell,+}|^2+|\psi_{\ell,-}|^2.
\label{eq:nl_def_SI}
\end{equation}
Then we have
\begin{equation}
|\psi_{\ell,+}|^2-|\psi_{\ell,-}|^2
=
n_\ell \sigma_\ell.
\label{eq:density_difference_SI}
\end{equation}
If the link intensities are approximately uniform,
\begin{equation}
n_\ell \approx n_0,
\label{eq:uniform_density_SI}
\end{equation}
then the polarization imbalance on each link is fully characterized by the Ising variable $\sigma_\ell$.

Assuming that each link condenses into a nearly pure circular polarization state with approximately fixed amplitude and phase reference, the complex polarization difference may be approximated by a fixed amplitude times the Ising sign, $\psi_{\ell,+}-\psi_{\ell,-}\approx A\sigma_\ell$. Accordingly, we have
\begin{equation}
\xi_v
=
\sum_{\ell\in v}\eta_{\ell v}(\psi_{\ell,+}-\psi_{\ell,-})
\approx
A\sum_{\ell\in v}\eta_{\ell v}\sigma_\ell,
\label{eq:Xi_to_Q_SI}
\end{equation}
where $A$ is an approximately constant amplitude factor set by the link occupation. This motivates the definition of the vertex charge
\begin{equation}
Q_v
=
\sum_{\ell\in v}\eta_{\ell v}\sigma_\ell.
\label{eq:Qv_def_SI}
\end{equation}
Equation~(\ref{eq:Xi_to_Q_SI}) then implies
\begin{equation}
|\xi_v|^2
\approx
A^2 Q_v^2.
\label{eq:Xi2_Q2_SI}
\end{equation}
Substituting Eq.~(\ref{eq:Xi2_Q2_SI}) into Eq.~(\ref{eq:Feff}) yields
\begin{equation}
F_{\mathrm{eff}}
\approx
U_{\mathrm{eff}}A^2
\sum_v Q_v^2.
\end{equation}
Defining
\begin{equation}
U = 2U_{\mathrm{eff}}A^2,
\end{equation}
we obtain the standard spin ice form
\begin{equation}
H_{\mathrm{ice}}
=
\frac{U}{2}\sum_v Q_v^2.
\label{eq:Hice_final_SI}
\end{equation}

For a four-coordinated vertex, each $\sigma_\ell=\pm1$, hence
\begin{equation}
Q_v \in \{-4,-2,0,+2,+4\}.
\end{equation}
The minimum of Eq.~(\ref{eq:Hice_final_SI}) is reached when
\begin{equation}
Q_v=0.
\label{eq:ice_rule_SI}
\end{equation}
This condition means that the number of incoming and outgoing Ising variables is equal at each vertex, i.e., the two-in two-out rule.

\begin{figure}
    \centering
    \includegraphics[width=1\linewidth]{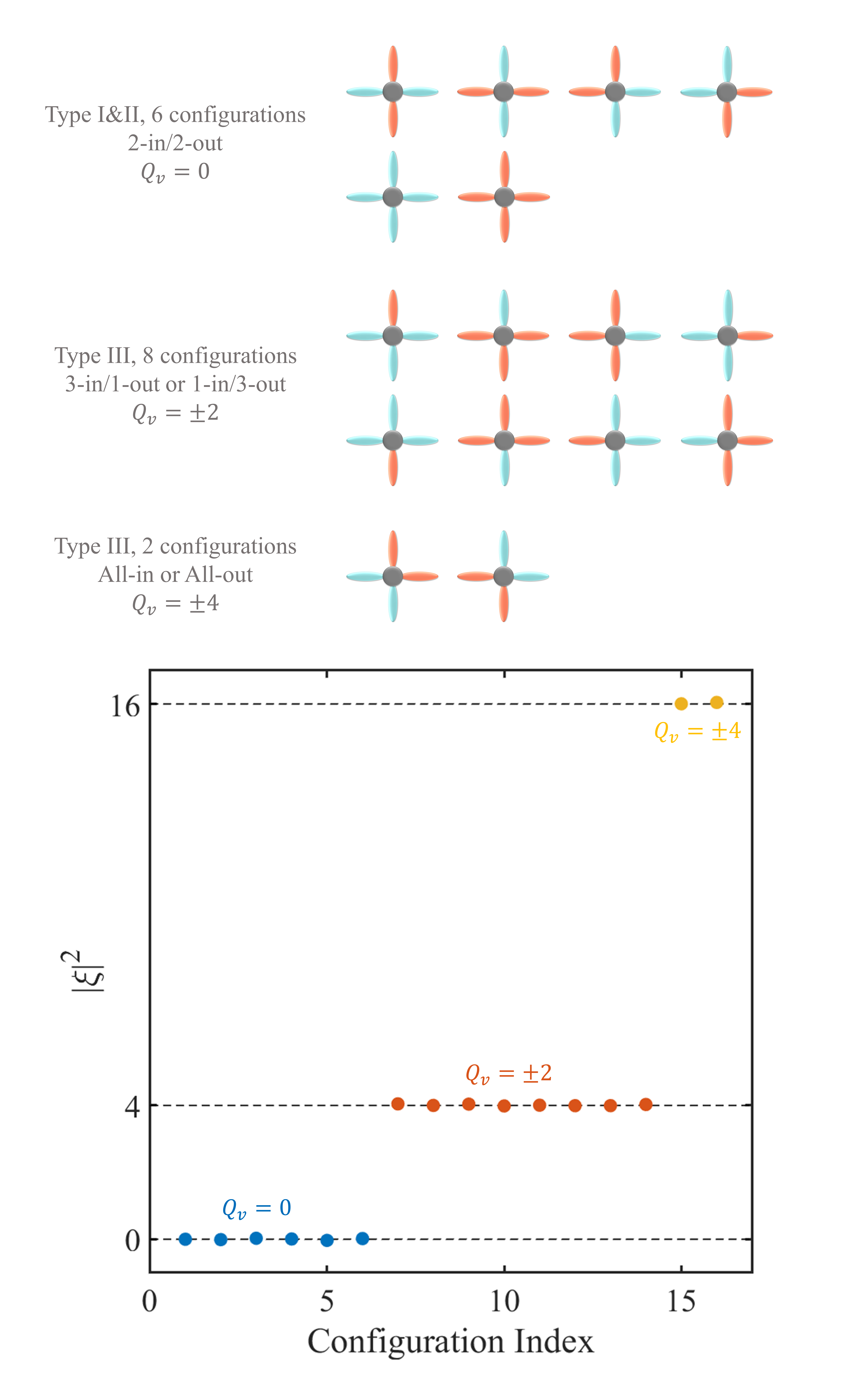}
    \caption{Total $2^4$ configurations of the artificial polariton spin ice unit and corresponding effective penalty energy $F_{\rm eff} = U_{\rm eff} |\xi|^2$.}
    \label{fig:Configurations}
\end{figure}

Configurations with $Q_v=\pm2$ correspond to three-in one-out or one-in three-out defects, while $Q_v=\pm4$ corresponds to all-in or all-out configurations.  Because the effective energy grows as $Q_v^2$, such defects are suppressed in the steady state unless they are created intentionally by local polarization flips. The $2^4$ configurations of a vertex with neighboring 4 links are shown in Fig.~\ref{fig:Configurations}. The energy of different configurations is measured by the energy of the stationary state corresponding to each configuration using Eqs. $6-7$. We further normalize the energy difference between $Q_v=0$ and $Q_v=\pm2$ states as 4. Consequently, the energy of all-in or all-out configurations spontaneously becomes 16.

\section{Experimental implementation of the lattice gauge structure and the vertex}

The central microscopic ingredient of our model is the oriented coupling
\begin{equation}
H_{\rm int}
=
g\sum_v
\left[
c_v^\ast
\sum_{\ell\in v}
\eta_{\ell v}\,
(\psi_{\ell,+}-\psi_{\ell,-})
+\mathrm{h.c.}
\right],
\label{eq:Hint_appD}
\end{equation}
where $\eta_{\ell v}=\pm1$ is the sign factor entering the vertex charge
\begin{equation}
Q_v=\sum_{\ell\in v}\eta_{\ell v}\sigma_\ell.
\end{equation}
For a realistic spin ice implementation, it is important that this sign structure be generated by the vertex-link coupling. We propose an experimental regime as following: each link retains an approximately degenerate pseudospin doublet, while the oriented sign $\eta_{\ell v}$ is engineered at the junction through the spatial parity and phase structure of the localized vertex mode. In this way, the low-energy selection is governed primarily by the vertex penalty proportional to $Q_v^2$.

Each lattice link $\ell$ may be realized as a quasi-one-dimensional polariton waveguide or an elongated micropillar supporting two circular polarization components $\psi_{\ell,+}$ and $\psi_{\ell,-}$. The key requirement is that these two components remain nearly degenerate, up to weak residual splittings that are much smaller than the effective vertex-induced scale. In practice, this means that the link geometry should avoid introducing a strong static polarization preference that would uniquely pin the pseudospin orientation on every bond. Experimentally, this may be achieved by using weakly anisotropic etched waveguides or micropillar links whose TE-TM splitting and birefringence remain small compared to the nonlinear gain competition and the vertex-induced dissipative scale. In this regime, the two circular components form an effective low-energy pseudospin $\pm1$ doublet, while the strong gain competition drives each link toward a nearly fully polarized state, $|\sigma_\ell|\approx 1$, without imposing in advance which of the two orientations would be chosen. This is the regime assumed in the Ising reduction discussed in Sec. III.

To realize the sign factor $\eta_{\ell v}$ microscopically, each four-coordinated junction should contain a localized auxiliary polariton mode $c_v$ whose overlap with the four surrounding links carries a controlled pattern of relative signs. The most natural way to obtain this is to use a structured vertex mode with nontrivial spatial parity. A convenient realization is a central cavity island or micropillar that supports a pair of low-lying odd-parity orbitals, such as $p_x$- and $p_y$-like confined modes \cite{barrat2024qubit,cherbunin2025quantum}. Because these modes change sign across the junction center, their overlap with opposite links naturally acquires opposite phases. This immediately generates an oriented sign structure of the form
\begin{equation}
(\eta_{L,v},\eta_{R,v},\eta_{D,v},\eta_{U,v})
=
(+1,-1,+1,-1),
\end{equation}
or any equivalent choice of the lattice gauge.

To implement such a structured lossy vertex, one may place a slightly larger cavity island at each lattice vertex, chosen such that its first excited confined modes are well separated and accessible. By tuning the pump and detuning, one can select one odd-parity vertex mode, or an appropriate superposition of two near-degenerate odd-parity modes, as the relevant auxiliary degree of freedom. The four adjacent waveguides are attached to the left, right, upper, and lower sides of the island, so that the coupling sign is inherited from the local parity of the confined field.

Alternatively, one can also engineer the contact region between each link and the vertex such that the effective tunnel matrix element acquires a controlled phase, $g_{\ell v}=g e^{i\phi_{\ell v}}$, with $\phi_{\ell v}=0$ or $\pi$. In the Ising setting only the sign matters, so this again reduces to $\eta_{\ell v}=\pm1$. This implementation is less geometrically transparent than the odd-parity orbital scheme, but may be useful if one wishes to realize the sign structure using patterned birefringent elements or polarization-dependent overlap regions.

In addition to the oriented sign, the vertex must couple to the difference of the two circular components on each link, namely $(\psi_{\ell,+}-\psi_{\ell,-})$. Experimentally, this can be realized by designing the junction mode to couple preferentially to a linearly polarized combination of the local spinor field, which in the circular basis becomes a relative-sign projection between $\psi_{\ell,+}$ and $\psi_{\ell,-}$. Equivalently, one may regard the contact region as defining an active pseudospin channel whose overlap with the vertex mode changes sign between the two circular components. Weak polarization anisotropy at the junction is sufficient for this purpose, provided that it acts primarily at the coupling interface and does not create a large on-link polarization splitting. In other words, the role of the junction anisotropy is to define how the link pseudospin is read out by the vertex, not to determine the preferred pseudospin state of an isolated link.

Finally, the link pseudospins can be reconstructed from polarization-resolved imaging of the emitted light. Since the sign convention is fixed by the vertex design, the local charges
\begin{equation}
Q_v=\sum_{\ell\in v}\eta_{\ell v}\sigma_\ell
\end{equation}
can be computed directly from measured polarization textures. Local polarization flips induced by resonant optical pulses or patterned pumps then create pairs of $Q_v=\pm2$ defects, allowing direct observation of monopole creation, motion, and annihilation.

\bibliography{apssamp}